\documentclass[fleqn,usenatbib]{mnras}
\usepackage{pdflscape}
\usepackage{float,lscape}
\usepackage{float}

\usepackage[T1]{fontenc}
\usepackage{ae,aecompl}
\usepackage{longtable}
\usepackage[table, dvipsnames]{xcolor}
\usepackage{comment}

 \usepackage{orcidlink}
\usepackage{newtxtext,newtxmath}

\usepackage[graphicx]{realboxes}
\usepackage{graphicx}	

\usepackage{tikz}
\usepackage{pgfplots}
\pgfplotsset{compat=1.14}
 \DeclareGraphicsExtensions{.png,.pdf}

\newcommand{\msun}{\,$M_{\odot}$}	
\definecolor{aliceblue}{rgb}{0.94,0.97,1.0}
\definecolor{grey}{RGB}{224,224,224}
\definecolor{mycolor2}{RGB}{199,233,252}
\definecolor{mycolor}{RGB}{255,204,204}

\title[GRB 210704A]
{Deciphering the unusual stellar progenitor of GRB~210704A}

\author[R.~L.~Becerra et al.]{R.~L.~Becerra\,\orcidlink{0000-0002-0216-3415}$^{1}$\thanks{E-mail: rosa.becerra@correo.nucleares.unam.mx (RLB)} 
E.~Troja\,\orcidlink{0000-0002-1869-7817},$^{2}$ 
A.~M.~Watson\,\orcidlink{0000-0002-2008-6927},$^{3}$
B.~O'Connor\,\orcidlink{0000-0002-9700-0036},$^{4,5,6}$
P.~Veres\,\orcidlink{0000-0002-2149-9846},$^{7,8}$ \newauthor
S.~Dichiara\,\orcidlink{0000-0001-6849-1270},$^{9}$
N.~R.~Butler\,\orcidlink{0000-0002-9110-6673},$^{10}$
F.~De~Colle\,\orcidlink{0000-0002-3137-4633},$^{1}$
T.~ Sakamoto\,\orcidlink{0000-0001-6276-6616},$^{11}$
K.~O.~C.~L\'opez\,\orcidlink{0000-0002-9322-6900},$^{3}$\newauthor
K.~Aoki\,\orcidlink{0000-0003-4569-1098},$^{12}$
N.~Fraija\,\orcidlink{0000-0002-0173-6453},$^{3}$
M.~Im\,\orcidlink{0000-0002-8537-6714},$^{13}$
A.~S.~Kutyrev\,\orcidlink{0000-0002-2715-8460},$^{4,5}$
W.~H.~Lee\,\orcidlink{0000-0002-2467-5673},$^{3}$\newauthor
G.~S.~H.~Paek\,\orcidlink{0000-0002-6639-6533},$^{13}$
M.~Pereyra\,\orcidlink{0000-0001-6148-6532},$^{14}$
S.~Ravi,$^{10}$ 
and
Y.~Urata\,\orcidlink{0000-0001-7082-6009}$^{15,16}$
\\
$^1$ Instituto de Ciencias Nucleares, Universidad Nacional Aut\'onoma de M\'exico, Apartado Postal 70-264, 04510 M\'exico, CDMX, Mexico\\
$^2$ Department of Physics, 
University of Rome - Tor Vergata, 
via della Ricerca Scientifica 1, 00100 Rome, IT\\
$^3$ Instituto de Astronom{\'\i}a, Universidad Nacional Aut\'onoma de M\'exico, Apartado Postal 70-264, 04510 M\'exico, CDMX, Mexico\\
$^4$ Department of Astronomy, University of Maryland, College Park, MD 20742-4111, USA\\
$^5$ Astrophysics Science Division, NASA Goddard Space Flight Center, 8800 Greenbelt Road, Greenbelt, MD 20771, USA\\
$^6$ Department of Physics, The George Washington University, 725 21st Street NW, Washington, DC 20052, USA\\
$^7$ Department of Space Science, University of Alabama in Huntsville, Huntsville, AL 35899, USA\\
$^8$ Center for Space Plasma and Aeronomic Research (CSPAR)\\
$^9$ Department of Astronomy and Astrophysics, The Pennsylvania State University, 525 Davey Lab, University Park, PA 16802, USA \\
$^{10}$ School of Earth and Space Exploration, Arizona State University, Tempe, AZ 85287, USA\\
$^{11}$ College of Science and Engineering, Department of Physics and Mathematics, Aoyama Gakuin University, 5-10-1 Fuchinobe, Chuo-ku,\\ Sagamihara-shi Kanagawa 252-5258, Japan\\
$^{12}$ Subaru Telescope, National Astronomical Observatory of Japan, 650 North A’ohoku Place, Hilo, HI 96720, USA\\
$^{13}$ Center for the Exploration of the Origin of the Universe (CEOU), Department of Physics and Astronomy, Seoul National University,\\ Seoul, 151-747, Republic of Korea\\
$^{14}$ CONACYT, Instituto de Astronom{\'\i}a, Universidad Nacional Aut\'onoma de M\'exico, 22860 Ensenada, BC, Mexico\\
$^{15}$ Institute of Astronomy, National Central University, Chung-Li 32054, Taiwan\\
$^{16}$ MITOS Science CO., LTD., New Taipei 235, Taiwan
}

\pubyear{2022}

\begin{document}
\label{firstpage}
\pagerange{\pageref{firstpage}--\pageref{lastpage}}
\maketitle

\begin{abstract}

GRB~210704A is a burst of intermediate duration ($T_{90} \sim 1-4$~s) followed by a fading afterglow and an optical excess that peaked about 7 days after the explosion.  
Its properties, and in particular those of the excess, do not easily fit into the well established classification scheme of GRBs as being long or short,
leaving the nature of its progenitor uncertain. 
We present multi-wavelength observations of the GRB and its counterpart, observed up to 160 days after the burst. 
In order to decipher the nature of the progenitor system, we present a detailed analysis of the GRB high-energy properties (duration, spectral lag, and Amati correlation), its environment, and late-time optical excess. 
We discuss three possible scenarios: a neutron star merger, a collapsing massive star, and an atypical explosion possibly hosted in a cluster of galaxies. We find that traditional kilonova and supernova models do not match well the properties of the optical excess, leaving us with the intriguing suggestion that this event was an exotic high-energy merger. 

\end{abstract}


\section{INTRODUCTION}
\label{sec:introduction}

Gamma-ray bursts (GRBs) are the brightest explosive events in the universe \citep{OConnor2023,Williams2023}. They are classified according to their duration $T_{90}$ and spectral hardness \citep{Kouveliotou93}. The population of short GRBs (SGRBs) typically has $T_{90} \lesssim 2$ s and harder spectra, whereas the population of long GRBs (LGRBs) typically has $T_{90} \gtrsim 2$ s and softer spectra \citep{Gehrels2013}. However, there is overlap between the two populations, and so in some cases it is not clear whether a burst with intermediate properties belongs to the one population or the other \citep{Garcia-Cifuentes2023,Troja2022,Yang2022,Dimple2022,Ahumada2021,Zhang2021,Gehrels2006}.

Nowadays, it is known that at least some SGRBs are the consequence of mergers between compact objects, driven by angular momentum and energy losses to gravitational radiation \citep[e.g.][]{Eichler1989,Narayan1992,Ruffert1998,Rosswog2002,Giacomazzo2011,Lee2007} and as such are sources of gravitational wave emission \citep{Abbott2017}. These mergers are followed by a luminous and short-lived kilonova emission, visible at optical and near-infrared wavelengths \citep{OConnor2021,Rastinejad2021,Troja2019,Drout17,Evans17,Tanvir13}.

In contrast, LGRBs are thought to be the result of the core-collapse of a star \citep{Woosley1993,MacFadyen1999ApJ,Hjorth2012} whose mass exceeds about $10 M_{\odot}$ \citep[see][for a review]{Woosley2006}.  
In this scenario, the optical emission from the SN appears a few days after the GRB, when the afterglow has faded sufficiently and the component powered by radioactive heating can be seen. This leads to two ways to identify the presence of a SN associated with a LGRB. First, through a “rebrightening” or “excess” in the optical /nIR light curve of the LGRB, visible for several days/weeks after the burst as a consequence of the SN emission \citep{Woosley1993,Galama1998,Becerra2017}. 
Second, by the appearance of the broad spectral lines characteristic of rapidly expanding ejecta.
The identification of a SN associated with GRB 980425  \citep{Galama1998} and of several more in the following years (see, e.g., \citealt{Hjorth2012}) confirmed that this scenario applies to most LGRBs. 

In this work, we investigate the nature of GRB 210704A, a burst first classified as a SGRB, but later considered to be of intermediate duration. We study the high-energy properties of the GRB prompt phase, and present  X-ray, optical, and infrared observations of its afterglow to 15 days after the burst. These observations show that the afterglow initially fades but then becomes brighter again in the optical and, to a less extent, in the near-infrared.  We also present late {\itshape HST} observations of a possible host galaxy. We derive constraints on the redshift of the bursts and consider three possible host environments: a nearby galaxy, a nearby cluster of galaxies, and a distant galaxy. We explore different scenarios to explain our data, and eliminate all of the those involving standard SGRBs or LGRBs. We are left with the intriguing possibility that GRB 210704A was an exotic transient in a cluster at $z\approx 0.2$.

Our paper is organised as follows. In section~\ref{sec:observations}, we present the observations with {\itshape Fermi}, {\itshape Swift}, {\itshape Chandra}, GTC, LDT, Gemini, Subaru, {\itshape HST}, and other telescopes. In section~\ref{sec:analysis} we present our analysis. We discuss the nature of the GRB in section~\ref{sec:discussion}. Finally, in section~\ref{sec:summary} we summarise our results. 

\section{Observations}
\label{sec:observations}

\subsection{Gamma-rays}

The prompt gamma-ray emission of GRB~210704A was detected by six instruments: {\itshape Fermi}/GBM \citep{30369,30380,30452}, {\itshape AGILE}/MCAL \citep{30372}, {\itshape Fermi}/LAT \citep{30375},  {\itshape AstroSat}/CZTI \citep{30378}, Konus-{\itshape Wind} \citep{30388}, {\itshape INTEGRAL}/SPI-ACS, and {\itshape INTEGRAL}/ISGRI \citep{30444}.
Hereafter, we use the {\itshape Fermi}/GBM trigger time as reference time  $T = $ 2021 July 4 19:33:24.59 UTC. 

\cite{30372} estimated a duration of about $T_{90} = 1.06$~s based on the  {\itshape AGILE}/MCAL data (0.4-100~MeV), and classified it as a SGRB. 
A similar short duration, $T_{90} = 1.0\pm 0.4$~s (20-200 keV), was derived by the analysis of the {\itshape AstroSat}/CZTI dataset \citep{30378}. 

\cite{30380} reported instead that the {\itshape Fermi}/GBM light curve consisted of a main bright peak with a duration $T_{90}$ of about 4.7~s (50-300~keV), with possible faint emission extending to $T+20$~s \citep{30452}. 
In the time interval between $T$ and $T+10$~s, high-energy ($>$100~MeV) emission was detected by the {\itshape Fermi}/LAT \citep{30375} with a photon flux of $(1.6\pm0.3)\times10^{-3}\ \mathrm{ph\,cm^{-2}\,s^{-1}}$ and a photon index of $-1.74\pm0.13$. 
The LAT localization was RA, Dec (J2000) = 159.08, +57.31 with a 90\% error radius of 6.7 arcmin, critical to enable follow-up observations with narrow-field instruments.

\cite{30388} reported Konus-{\itshape Wind} observations of a bright peak
in the interval from $T$ to $T+2$~s followed by a weaker pulse peaked at $T+4.5$~s (20~keV - 4~MeV). Finally, \cite{30444} estimated a duration of $T_{90} = 3.5\pm 0.7$~s ($>$80~keV) based on the {\itshape INTEGRAL}/SPI-ACS data. We noted that {\itshape INTEGRAL}/ISGRI also observed the event but unfortunately observations were saturated \citep{30444}.
All the durations $T_{90}$ cited above are summarised in Table~\ref{tab:durations}. 

\begin{table}
	\centering
	\caption{Duration $T_{90}$ reported by different facilities for GRB 210704A}
	\label{tab:durations}
 \begin{tabular}{lccc} 
		\hline
			Instrument & $T_{90}$ [s] & Energy [keV]& Reference\\
		\hline
    
{\itshape AGILE}/MCAL&1.06&400-100000& 1\\
{\itshape AstroSat}/CZTI&4.6$^{+2.8}_{-3.6}$&20-200& 2\\
{\itshape Fermi}/GBM &4.7 &50-300& 3 \\
Konus-{\itshape Wind}&4.5&20-4000& 4\\
{\itshape INTEGRAL}/SPI-ACS&$3.5\pm 0.7$&80-1000& 5 \\

		\hline\\
	\end{tabular}
References: (1) \cite{30372}; (2) Private communication; 
(3) \cite{30380}; (4) \cite{30388}; (5) \cite{30444}. 
\end{table}

\begin{figure}
\centering
 \includegraphics[width=\columnwidth]{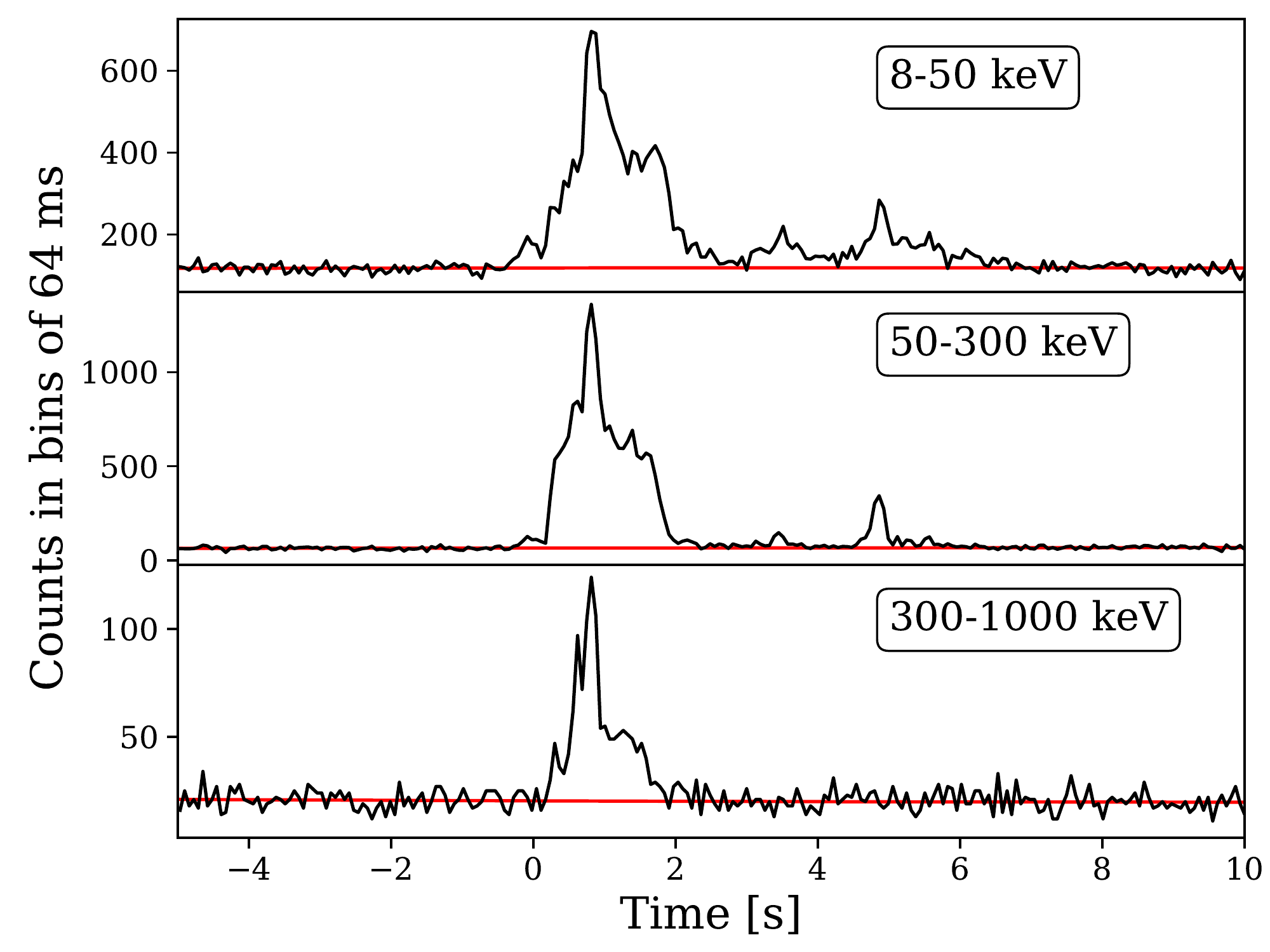}
  \vspace{-0.5cm}
  \caption{The {\itshape Fermi}/GBM gamma-ray light curve (black) and background (red) of GRB 210704A in 3 energy bands. Note the absence of the extended emission in the hardest band.}
 \label{fig:gammalc2}
\end{figure}

The values of $T_{90}$ reported in GCNs range from 1.0 to 4.7 seconds (see Table~\ref{tab:durations}) and do not unambiguously aid in the classification of GRB~210704A. 
This discrepancy can largely be understood by considering Figure~\ref{fig:gammalc2}, which shows the Fermi/GBM light curves in three energy bands.
It shows that the observed emission in the soft energy range (8-50~keV; top panel) consists of a weak precursor, followed by strong initial peak with a duration of less than 2 seconds and then an extended tail out to about 6 seconds with two weaker peaks at about 3.5 and 4.8 seconds. 
However, the precursor and tail are not seen in the harder 300-1000~keV band. 
Therefore, we suggest that gamma-ray instruments that are not sensitive to soft ($\lesssim$50 keV) energies would not detect the extended emission and would measure shorter values of $T_{90}$.  
This sort of instrumental selection effect is well known \citep{Qin2013,Moss2022} and can explain the short $T_{90}$ measured by {\itshape AGILE}/MCAL, but does not completely explain the short value measured by {\itshape AstroSat}/CZTI. 

We conclude that from the prompt gamma-ray light curve alone it is not clear whether the GRB is a LGRB or a SGRB with temporally extended spectrally soft emission.

\begin{table*}
	\centering
	\caption{Photometry of GRB 210704A.} 
	\label{tab:observations}
 \begin{tabular}{lccrlr} 
		\hline
			Time [d] & Exp [s]& Filter & Magnitude & Instrument &Reference \\
		\hline
    0.38 & 360& \emph{ w }& $>$ 19.70 & DDOTI & \cite{30383}, This work \\
0.73 & 3038& \emph{ v }& $>$ 21.10 & UVOT/SWIFT & \cite{30389} \\
0.94 & 3960& \emph{ r' }& 22.25 $\pm$ 0.13 & AZT-20 & \cite{30384} \\
1.08 & 1560& \emph{ r }& 22.13 $\pm$ 0.13 & DOLORES/TNG & \cite{30385} \\
1.08 & 900& \emph{ z' }& $>$ 21.50 & OAJ-T80 & \cite{30401} \\
1.09 & 900& \emph{ i' }& 22.08 $\pm$ 0.23 & OAJ-T80 & \cite{30401} \\
1.09 & 1500& \emph{ i' }& 21.84 $\pm$ 0.13 & CAFOS/CAHA & \cite{30391} \\
1.09 & 120& \emph{ r }& 22.40 $\pm$ 0.07 & OSIRIS/GTC & \cite{30392} \\
1.10 & 900& \emph{ g' }& 22.42 $\pm$ 0.11 & OAJ-T80 & \cite{30401} \\
1.12 & 900& \emph{ r' }& 22.17 $\pm$ 0.10 & OAJ-T80 & \cite{30401} \\
1.29 & 1152& \emph{ i }& $>$ 22.10 & RATIR & \cite{30390} \\
1.29 & 1152& \emph{ r }& $>$ 22.40 & RATIR & \cite{30390} \\
2.07 & 27000& \emph{ r }& $>$ 22.70 & CAFOS/CAHA & \cite{30411} \\
2.10 & 420& \emph{ r }& 23.43 $\pm$ 0.11 & OSIRIS/GTC & This work \\
2.10 & 420& \emph{ z }& 23.44 $\pm$ 0.30 & OSIRIS/GTC & This work \\
3.92 & 4620& \emph{ r' }& $>$ 23.10 & AZT-20 & \cite{30440} \\
4.10 & 4440& \emph{ i }& 23.39 $\pm$ 0.14 & DOLORES/TNG & \cite{30432} \\
4.10 & 4440& \emph{ r }& 23.35 $\pm$ 0.16 & DOLORES/TNG & \cite{30432} \\
4.45 & 840& \emph{ K }& 23.00 $\pm$ 0.30 & NIRI/Gemini-North & This work \\
5.10 & 1020& \emph{ r }& 23.68 $\pm$ 0.10 & OSIRIS/GTC & This work \\
5.11 & 1020& \emph{ z }& 22.56 $\pm$ 0.15 & OSIRIS/GTC & This work \\
5.50 & 1440& \emph{ J }& 22.20 $\pm$ 0.30 & NIRI/Gemini-North & This work \\
6.08 & 900& \emph{ g }& 23.78 $\pm$ 0.11 & AlFOSC/NOT & \cite{30443} \\
6.10 & 900& \emph{ r }& 23.27 $\pm$ 0.09 & AlFOSC/NOT & \cite{30443} \\
6.35 & 1500& \emph{ r }& 23.14 $\pm$ 0.27 & LMI/LDT & This work \\
6.45 & 2580& \emph{ K }& 23.10 $\pm$ 0.20  & NIRI/Gemini-North & This work \\
6.98 & 3600& \emph{ R }& 23.10 $\pm$ 0.30 & Zeiss-1000/SAO-RAS & \cite{30465} \\
10.44 & 1080& \emph{ r }& 23.73 $\pm$ 0.10 & GMOS-N/Gemini-North & This work \\
11.44 & 900& \emph{ z }& $>$ 22.44 & GMOS-N/Gemini-North & This work \\
12.45 & 1800 & \emph{ K }& 23.10 $\pm$ 0.20 & NIRI/Gemini-North & This work \\
14.45 & 1584& \emph{ z }& 23.58 $\pm$ 0.24 & GMOS-N/Gemini-North & This work \\
115.00 & 3030& \emph{ r }& $>$ 25.21 & HSC/Subaru & This work \\
161 & 7200 &\emph{ K } & $>24.0$ & NIRI/Gemini-North & This work \\

		\hline
	\end{tabular}
\end{table*}

\begin{table*}
	\centering
	\caption{Photometry of the  galaxies S1 and S2.} 
	\label{tab:gal_phot}
 \begin{tabular}{lccrl} 
		\hline
			Source & Exp [s]& Filter & Magnitude & Instrument \\
		\hline
    S1 & 2120 & \textit{F336W} & $26.10\pm0.30$ &WFC3/UVIS  \\
... & 2120 &  \textit{F438W} & $25.80\pm0.20$ & WFC3/UVIS \\
... & 1600 &  \textit{F606W} & $25.83\pm0.12$ & WFC3/UVIS  \\
... & 1020&\emph{r}  &25.30 $\pm 0.20$ & OSIRIS/GTC  \\
... & 1200 & \textit{F105W} & $25.61\pm0.14$ & WFC3/IR  \\
... & 1350 & \textit{F160W} & $24.46\pm0.08$  & WFC3/IR  \\
\hline
S2 & 2120 & \textit{F336W} & $25.65\pm0.17$ &WFC3/UVIS  \\
... & 2120 &  \textit{F438W} & $25.32\pm0.17$ & WFC3/UVIS \\
... & 1600 &  \textit{F606W} & $24.80\pm0.07$ & WFC3/UVIS  \\
... & 1020&\emph{r}  & $24.68\pm 0.08$ & OSIRIS/GTC  \\
... & 1200 & \textit{F105W} & $24.04\pm0.05$ & WFC3/IR  \\
... & 1350 & \textit{F160W} & $23.97\pm0.05$  & WFC3/IR  \\
		\hline
	\end{tabular}
\end{table*}

\begin{figure*}
\centering
 \includegraphics[width=0.95\textwidth]{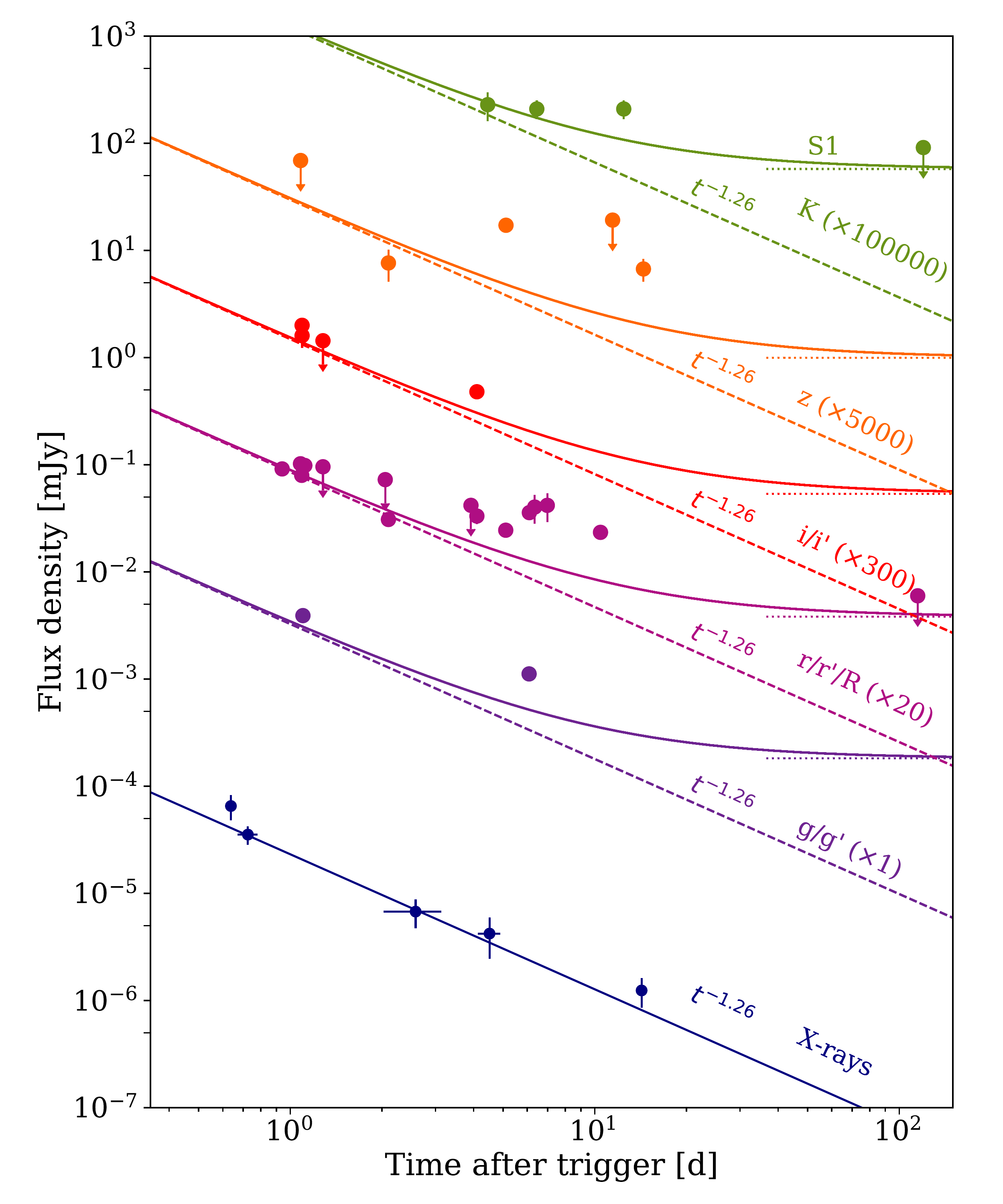}
 \caption{X-ray, optical and near-infrared light curves of GRB 210704A. Upper limits are at $3\sigma$. The dashed lines show the power-law fits to the afterglow, dotted horizontal line show the contribution of the underlying source S1 in each filter, and the solid lines show the combination of the two components.}
 \label{fig:lightcurve}
\end{figure*}

\subsection{X-rays}
\label{sec:x-rays}

Following the localisation by the {\itshape Fermi}/LAT,
the X-Ray Telescope (XRT) aboard {\itshape Swift} initiated automated Target of Opportunity (ToO) observations of the field \citep{30374}. 
A first visit was carried out between $T+53.5$~ks and  $T+61.6$~ks for a total exposure of 2.6~ks. This led to the identification of the X-ray afterglow at an enhanced position of RA, Dec (J2000) = 10:36:05.21 +57:12:59.1 with a 90\% uncertainty radius of 2.7\arcsec.
Monitoring of the source continued until $T + 5.3$~d, when the X-ray afterglow fell below the XRT detection threshold. 

In our analysis, we use light curves and spectra from the UK Swift Science Data Centre (UKSSDC) on-line repository \citep{Evans2009}. 
The XRT light curve displays a simple power-law decline, $t^{-\alpha}$ with $\alpha=1.3\pm0.2$.
The time-averaged XRT spectrum, from $T+53$~ks to $T+455$~ks, is best described by an absorbed power-law with a photon index $\Gamma=1.7\pm0.2$ and hydrogen column density $N_{\rm H}$ = $5.6\times 10^{19}\ \mathrm{cm^{-2}}$ fixed at the Galactic value. 
Based on this model, the unabsorbed X-ray flux (0.3–10 keV) at 11~hrs is approximately $8\times10^{-13}\ \mathrm{erg\,cm^{-2}\,s^{-1}}$, in agreement with the distribution of X-ray fluxes of bright SGRBs \citep{OConnor2020}.

In order to characterise the afterglow temporal evolution at late times, we requested Director's Discretionary Time (DDT) observations with the {\itshape Chandra} X-ray Telescope. The target was observed with the ACIS-S3 camera, starting at 01:32:42 UTC on 2021 July 22 ($T+14.25$~d) for a total exposure of 19.8~ks (ObsID: 25093; PI: Troja). The \textit{Chandra} data were processed using \texttt{CIAO} 4.12 with \texttt{CALDB} Version 4.9.0. 
At the afterglow position we detect 12 photons within a $1\arcsec$ extraction radius. 
After correcting for PSF-losses, we derive a count rate of $(6.6^{+2.1}_{-1.7}) \times 10^{-4}\ \mathrm{cts\,s^{-1}}$ in the 0.5-7.0~keV energy range.
This translates to an unabsorbed X-ray flux of $(1.3^{+0.4}_{-0.3}) \times 10^{-14}\ \mathrm{erg\,cm^{-2}\,s^{-1}}$ in the 0.3-10~keV energy band using the best fit parameters derived from the {\itshape Swift}/XRT spectrum. 
The combined {\itshape Swift}/{\itshape Chandra} light curve is best described by a power-law decay with $\alpha=1.26\pm0.04$.

\subsection{Optical and Infrared Photometry}
\label{sec: optical}

We carried out an extensive campaign of follow-up observations using 
the Deca-Degree Optical Transient Imager (DDOTI; \citealt{30383}),
the Large Monolithic Imager (LMI) on the 4.3-m Lowell Discovery Telescope (LDT; \citealt{30451}), the Near Infra-red Imager (NIRI) and the Gemini Multi-Object Spectrographs (GMOS) instruments on the 8.1-m Gemini-North Telescope  \citep{30442},  the Hyper Suprime-Cam (HSC) on the Subaru 8.2 m telescope \citep{Miyazaki2018}, and the Optical System for Imaging and low-Intermediate-Resolution Integrated Spectroscopy (OSIRIS) instrument on the 10.4-m Gran Telescopio de Canarias (GTC; \citealt{30436}). 
Table~\ref{tab:observations} gives a log of the observations. 
This dataset was supplemented by archival observations from the
\textit{Hubble Space Telescope} (\textit{HST}) obtained with the Wide Field Camera 3 (WFC3), and 
from the Canada-France-Hawaii Telescope (CFHT) obtained with MegaPrime/MegaCam during December 2004 (PI: Tanvir and Cowie) (observationID: G013.158.521+57.697) 
and retrieved from the MegaPipe image stacking pipeline\footnote{https://www.cadc-ccda.hia-iha.nrc-cnrc.gc.ca/en/megapipe/access/graph.html}. 

{We show the field of GRB 210704A observed with the \textit{HST} in Figure~\ref{fig:field}. We discuss in detail the host galaxy candidates of our event in \S~\ref{sec:host}.

\begin{figure*}
\centering
 \includegraphics[width=1.99\columnwidth]{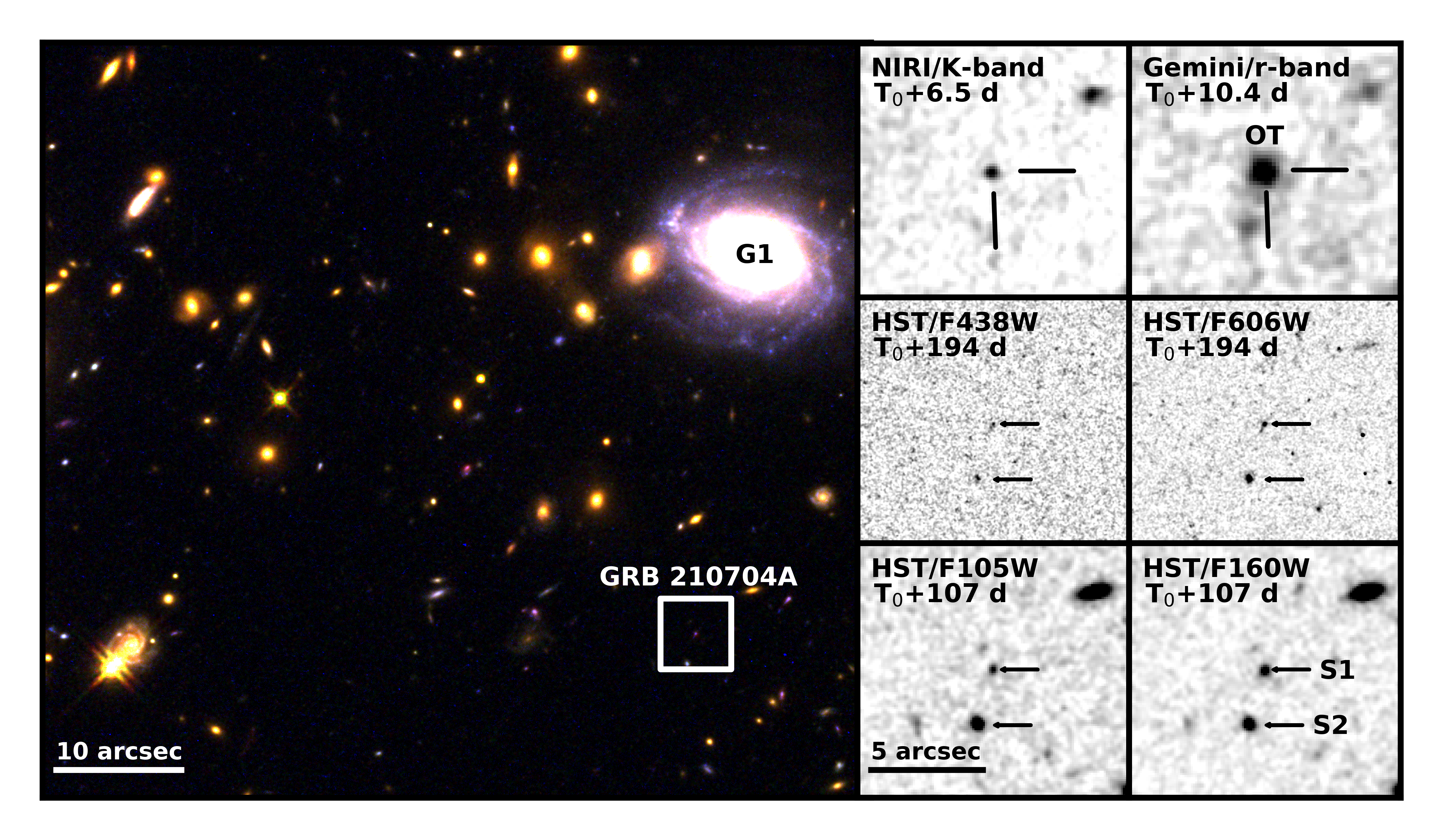}
 \caption{False-color image of the field of GRB 210704A as seen by \textit{HST}: red is $F160W$, green is $F105W$, and blue is $F606W$. The white box in the bottom right marks the position of GRB 210704A. The bright spiral galaxy G1 is marked in the top right. Side panels show a zoom of the field in different filters to compare the afterglow images obtained at $6.5$ and $10.4$ d with the late-time \textit{HST} imaging obtained months after the explosion. The crosshair in the top panels marks the optical/infrared transient position, and the arrows mark the location the sources S1 and S2, respectively. The orientation is such that North is up and East is to left. 
 }
  \label{fig:field}
\end{figure*}

Data were processed following standard techniques for CCD data reduction (e.g., bias subtraction, flat-fielding, cosmic-ray rejection).  
The Gemini images were reduced using the Gemini \texttt{IRAF} package v1.14\footnote{https://www.gemini.edu/observing/phase-iii/understanding-and-processing-data/data-processing-software/gemini-iraf-general}, whereas the GTC, LDT and Subaru images were reduced using custom Python-based software. The \textit{HST} data were processed using standard procedures within the \texttt{DrizzlePac} package \citep{Gonzaga2012}. The final pixel scale was $0.09\arcsec$/pix for WFC3/IR ($F105W$ and $F160W$ filters) and $0.03\arcsec$/pix for the WFC3/UVIS data.


We performed PSF photometry on the optical images using \texttt{SExtractor} \citep{Bertin1996}. To minimize systematic errors in the calibration, we calibrated all of the $r$ and $z$ photometry using the same six nearby Sloan Digital Sky Survey (SDSS) Data Release 12 (DR12) stars \citep{{SDSS12}}. 
In the nIR images from Gemini, only a weak signal is visible at the GRB position. We estimated its brightness by performing forced aperture photometry, calibrated against nearby Two Micron All-Sky Survey (2MASS; \citealt{Cutri2003}) sources. Due to the low number of 
2MASS sources within the NIRI field of view and their likely extended nature, a systematic error of 0.1 mag was added to the absolute calibration of the nIR photometry.  

We performed photometry of the two galaxies S1 and S2 (see Figure~\ref{fig:field}) in the late-epoch \textit{HST} images using using \texttt{SExtractor} aperture photometry and the photometric zeropoints in the image headers. We will discuss the possibility of these galaxies being the host of the GRB in \S\ref{sec:host}, but for the time being we will simply note that S1 is almost coincident with the GRB position and therefore our ground-based observations combine the flux of both the GRB counterpart and S1.

Our photometry is shown in Table~\ref{tab:observations} (with upper limits being $3\sigma$) and Table \ref{tab:gal_phot} and is plotted in Figure~\ref{fig:lightcurve}. All our magnitudes were converted to an AB scale and not corrected for Milky Way extinction, $E(B-V)$ = 0.007 \citep{Schlegel1998}. We also include additional measurements from the literature \citep{30384,30385,30389,30391,30392,30401,30411,30432,30440,30443,30465}. 


\subsection{Optical Spectroscopy}
\label{sec:opticalspec}
Optical spectroscopy of the afterglow at $T + 1.1$~d was obtained by \cite{30392} 
using the OSIRIS instrument on GTC. We downloaded their calibrated spectrum from the GRBSpec database \citep{deUgarte2014}, and show it in Figure~\ref{fig:spectrum}. 


\cite{30392} noted the clear presence of a continuum down to at least 4200~{\AA}, which implies $z < 3.6$. 
They also noted a tentative detection of a broad absorption dip at about 4050~{\AA}, which they interpreted as Ly-$\alpha$ at $z = 2.34$. We note that this feature occurs in a part of the spectrum that is quite noisy, which calls its reality into doubt.
They also commented on possible low-significance detections of absorption from \ion{O}{I} (1302/1304~{\AA}), \ion{Si}{II} (1260~{\AA}), and \ion{C}{II} (1334~{\AA}),
but the absence of corresponding absorption from \ion{C}{IV} and \ion{Si}{IV}.

Another clue to the GRB redshift comes
from the lack of \ion{Mg}{II} absorption features.
The \ion{Mg}{II} doublet was observed with a mean rest-frame equivalent width of about 3.8~{\AA} in 27 of the sample of 31 LGRBs observed by \cite{deUgarte2012} and \cite{Fynbo2009}. 
The optical spectrum of GRB210704A has a signal-to-noise ratio (SNR) of at least 5 from 5000 to 7200~{\AA} and shows no strong absorption lines. For this reason, we suggest that if the GRB is similar to typical LGRBs, then the absence of \ion{Mg}{II} disfavours the redshift range 
$0.8<z<1.6$. However, if the GRB is similar to typical SGRBs, whose spectra do not show strong host absorption lines, then we have the weaker constraint that the redshift is less than 3.6.

\begin{figure*}
\centering
 \includegraphics[width=0.95\textwidth]{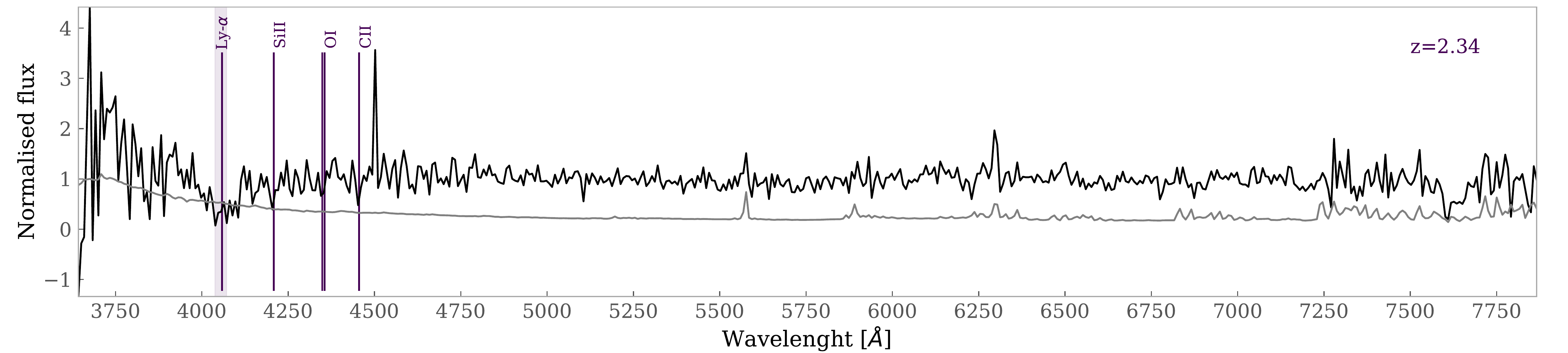}
 \caption{
 OSIRIS/GTC normalized optical spectrum (black line) and the corresponding error (grey line). The spectrum was obtained and reduced by \citep{30392} and  retrieved from the GRBSpec database. 
\citet{30392} noted possible Ly$\alpha$ absorption at z=2.34.} 
 \label{fig:spectrum}
\end{figure*}

\section{Analysis}
\label{sec:analysis}

\subsection{Amati correlation}
\label{sec:amati}

Given the ambiguity of the GRB duration, we will consider its properties in the context of the Amati correlation \citep{Amati2008}. For this, we analysed the {\itshape Fermi}/GBM data from $T-0.2$ to $T+6.1$~s and found the best fitting function was a Band function with $\alpha=-0.59\pm0.02$, $\beta=-2.95\pm 0.20$, and $E_{\rm peak}= 283\pm 8 $ keV. 
The fluence in the 10-1000 keV interval (observer's frame) is $F=(2.3\pm0.3)\times 10^{-5}$ erg cm$^{-2}$.  In the top panel of Figure~\ref{fig:amati-lag}, we show the position of the GRB at $z = 0.1$, 0.4, 0.8, and 2.34. We also show the populations of SGRBs and LGRBs from \cite{Amati2008}. We see that GRB 210704A is consistent with being a SGRB at low redshift ($z<0.4$) or with being a LGRB at higher redshift ($z \gtrsim 0.4$).

\subsection{Spectral lag}
\label{sec:lag}

We also consider the spectral lag GRB 210704A in the context of other GRBs, following \cite{Norris2000} and \cite{Gehrels2006}. 
We derive the GRB peak flux by fitting the spectrum of the brightest 1.024 s interval \citep{Poolakkil2021}  with a Band function.
The best fit model yields $\alpha=-0.36\pm 0.02$, $\beta=-2.82\pm 0.12 $, $E_{\rm peak}= 302\pm 8$ keV), and a peak flux of $F=(1.53\pm 0.02)\times 10^{-5}$ erg cm$^{-2}$ s$^{-1}$. 
From this value, we derive the isotropic-equivalent peak luminosity,
 $L_{\rm peak}$, by assuming different redshifts.
 For example, at $z=0.5$ the measured peak flux corresponds to $L_{\rm peak}  = (5.46 \pm 0.07)\times 10^{51}\ \mathrm{erg\,s^{-1}}$ in the comoving 50--300~keV energy range.

We calculated the lag between the energy bands 25--50 keV and 100--300 keV using the 10 ms resolution light curve and the sum of the signals from the three brightest {\itshape Fermi} \ion{Na}{I} detectors (1, 3 and 5). 
The lag was determined by the maximum of the cross-correlation function between the two light curves. We fit a 4th-degree polynomial around the peak in order to accurately determine the lag. We find that the lag is:
$\tau = 80\pm 9~{\rm ms}$,
in which the uncertainties are $1\sigma$.
We estimated the uncertainty by adding Poisson noise to the data \citep[for a similar approach, see e.g., ][]{Ukwatta+10lag,Hakkila+18smoke} and repeating the cross-correlation analysis.

In the bottom panel of Figure~\ref{fig:amati-lag}, we show the spectral lag (corrected for redshift as $(1+z)^{0.67}$, following \citealt{Gehrels2006}) and peak luminosity $L_{\rm peak}$ of the GRB at $z=0.1$, 0.4, 0.8, and 2.34. We also show the populations of LGRBs and SGRBs from \citep{Norris2000} and \cite{Gehrels2006}. 
We see that, for any redshift, the GRB lag is longer than typical SGRBs.
Its luminosity fits within the distribution
of cosmological LGRBs for $z \gtrsim 0.8$ (and also for $z \sim 0.4$ is marginally consistent with the rest of the LGRB, however, it is outside the region of 2$\sigma$ scatter), whereas for lower redshifts its location in the lag-luminosity diagram would be unusual. 
Other outliers in a similar position are 
GRB~031203 and GRB~060729, peculiar LGRBs associated with bright supernovae.

\begin{figure}
\centering
 \includegraphics[width=0.99\columnwidth]{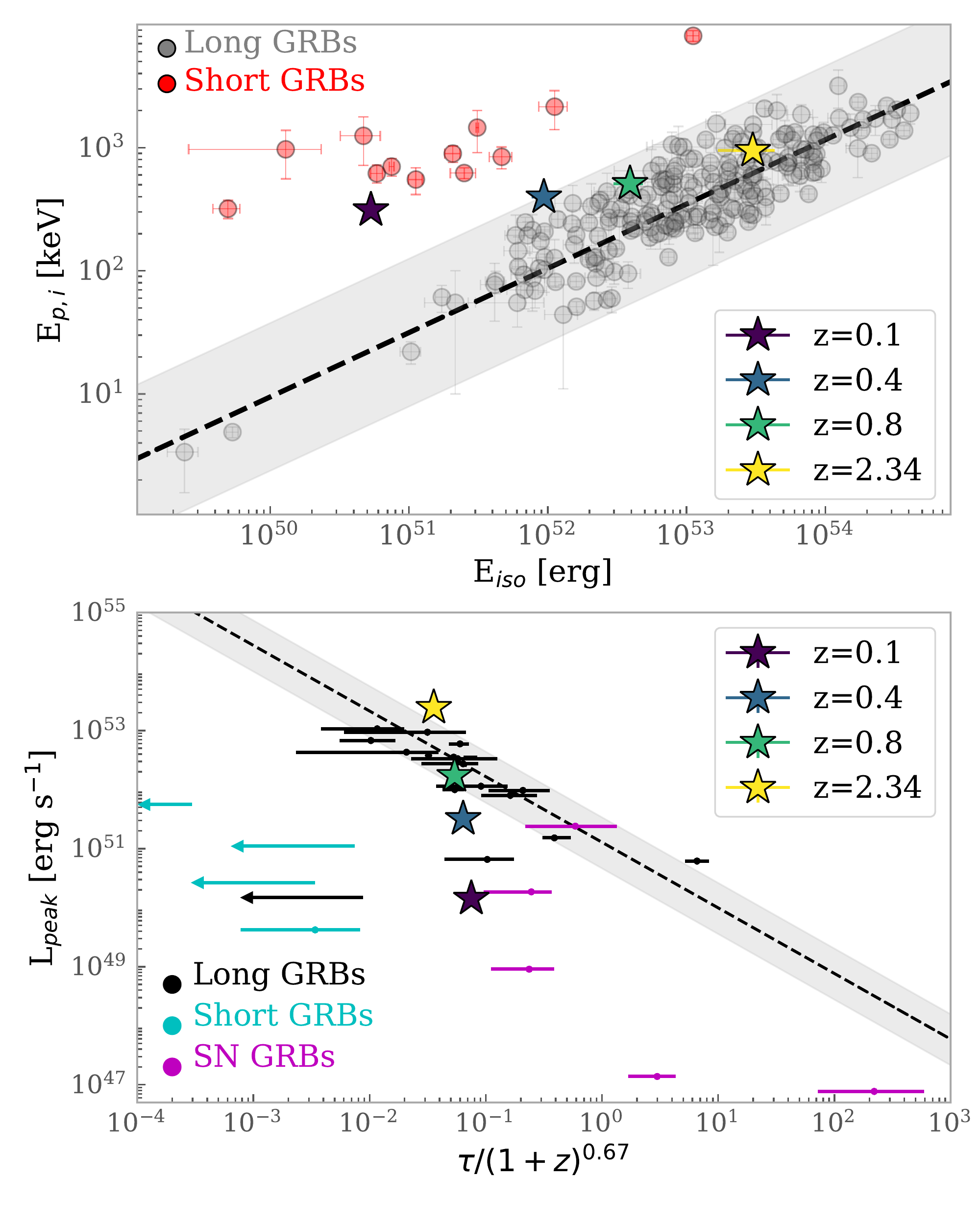}
 \caption{Top panel: The $E_{\rm peak}$ and $E_{\rm iso}$ Amati relation. Bottom panel: The $\tau$ and $L_{\rm peak}$ relations.  The black dotted lines are the relations from \protect \cite{Amati2008} and \protect\cite{Norris2000} and the grey regions show the 2$\sigma$ scatter.
 The lower figure also shows SGRBs from \protect\cite{Norris2000}, including the energy band correction. 
 Both panels show the estimated positions of GRB 210704A at $z=0.1$, 0.4, 0.8, and 2.34}

 \label{fig:amati-lag}
\end{figure}

\begin{figure}
\centering

 \includegraphics[width=0.95\linewidth]{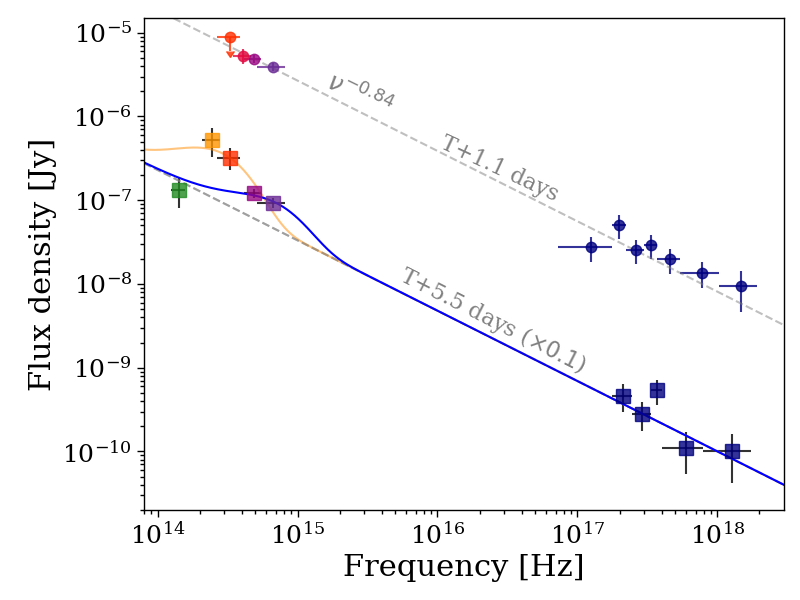}
 
 \caption{Optical and X-ray SED of GRB 210704A at $T+1.1$~d (circles) and $T+5.5$~d (squares). Upper limits are at $3\sigma$. The colours of the symbols are the same as in Figure~\ref{fig:lightcurve}. 
 The early epoch is well described by a power-law spectrum with slope 0.84 (dashed line). 
 The second epoch shows optical/nIR emission in excess of the simple power-law. This excess is modeled as a blackbody component with observed temperature between $T=3800$ K (orange thin line) and $T=9000$ K (blue thick line).
}
 \label{fig:sed}
\end{figure}

\subsection{Afterglow}
\label{sec:ag}

In the standard fireball model for GRBs, the afterglow emission is produced by external shocks resulting from the interaction between the relativistic jet and the circumstellar medium \cite[e.g.,][]{Kumar2000,Fraija2015}. 
{Typically, the afterglow phase can be explained as synchrotron radiation from shocked electrons $N(E)\propto E^{-p}$, which results in a series of power-law segments in the observed flux density as a function of time $t$ and frequency $\nu$. These power-law segments are characterized by temporal and spectral indices, denoted as $\alpha$ and $\beta$ respectively, and are described by the equation $F_{\nu}\propto t^{-\alpha}\nu^{-\beta}$ \citep{Sari1998,Granot2002}.

The early optical and X-ray data can be fit with a simple absorbed power-law of slope $\beta=0.84\pm0.02$ (see Figure~\ref{fig:sed}), indicating that they belong to the same spectral segment and therefore they are expected to display the same temporal decay. 
We identify this segment as $\nu_m\lesssim\nu_{opt}\lesssim\nu_{X}\lesssim\nu_c$ (and therefore a slow-cooling regime), where $\nu_m$ is the synchrotron characteristic frequency and $\nu_c$ is the cooling frequency \citep{Granot2002}. 
In this case, $F_\nu\propto \nu^{(1-p)/2}$ and the electron index is $p=2.68\pm0.04$.
For a fireball expanding into a uniform density medium, the relation of flux density and time is $F_\nu\propto t^{3(1-p)/4}$. With the electron index previously estimated, we obtain $F_\nu\propto t^{-1.26\pm0.03}$.


A simple power-law decay of slope -1.26 is consistent with X-ray observations at all epochs and with the early optical observations (see Figure~\ref{fig:lightcurve}). However, beginning at $T$+5.5~d, it underpredicts the observed optical and near-infrared emission. 
We interpret this as evidence of an additional component powering the late-time optical/nIR light curve and will return to discuss it in detail in \S\ref{sec:excess}.


From the condition that the X-ray afterglow is in slow-cooling regime, we derive an upper limit on the circumburst density $n$ following \citet{OConnor2020}: 
\begin{equation}
\begin{split}
n \lesssim  0.4 & \left(\frac{1+z}{{\rm 2}}\right)^{-6/11} 
\left(\frac{t}{{\rm 1~d}}\right)^{-4/11} 
\left(\frac{E_{\rm K,iso}}{10^{52} {\rm~erg}}\right)^{-6/11}\\
& \left(\frac{\epsilon_e}{0.1}\right)^{-40/33}
\left(\frac{\epsilon_B}{10^{-5}}\right)^{-5/9}
{\rm~cm^{-3},}
\end{split}
\end{equation}
which is consistent with densities typical of the interstellar medium, and disfavors the high-density environment where some LGRBs are found.

Finally, by assuming that the GeV flux is also produced by external forward shocks, 
we can use the flux above 100~MeV \citep{30375} as proxy for the blastwave kinetic energy \citep{Beniamini2015}. For $p\approx2.6$, we obtain: 
\begin{equation}
\begin{split}
 E_{\rm K,iso}  \approx &~  10^{53}  \left(\frac{1+z}{3}\right)^{-1} \left(\frac{d_{L,28}}{5}\right)^{1.74}
\left(\frac{F_{LAT}}{{\rm 0.2~nJy}}\right)^{0.87}
\left(\frac{t}{{\rm 5~s}}\right)^{1.25} \\
&  \left(\frac{\epsilon_e}{0.1}\right)^{-1.38}
\left(\frac{\epsilon_B}{10^{-3}}\right)^{0.13}
{\rm~erg,}
\end{split}
\end{equation}
where $d_{L,28}$ is the luminosity distance in units of $10^{28}$~cm.
For $z\gtrsim$2, this value fits well within the distribution of LGRBs.

\subsection{Late-time Excess}
\label{sec:excess}

The light curves in Figure~\ref{fig:lightcurve} are compared to the standard power-law afterglow.  A late-time flattening is expected due to the contribution of the source S1, underlying the GRB position. This model (solid lines) underpredicts the optical and, to a less extent, near-infrared photometry from $T+5$~d to $T+15$~d. We consider this to be indicative of an excess component in addition to the afterglow. We identify the peak of this excess at around $T+6.5$~d, when the observed light is 3-5 times brighter than the afterglow+S1 model. 

Figure~\ref{fig:sed} shows the spectral energy distribution (SED) of the GRB counterpart at 
two epochs, $T+1.1$~d and  $T+5.5$~d, respectively. 
The earlier SED is well described by a power-law function, as expected for a non-thermal afterglow spectrum.
The later SED shows an excess in the optical ($grz$) and $J$ band. 
We consider a simple model for the excess, treating it as black-body emission arising from a spherical fireball in expansion and ignoring relativistic effects. The rest-frame parameters of this model are the temperature $T'$ and radius $R'$ of the photosphere. 
In the observer's frame, the spectrum will have the shape of a black-body with $T_{\rm obs} = T'/(1+z)$.
If we fit this simple black-body model to the $grzJK$ data in Figure~\ref{fig:sed}, we obtain $T_{\rm obs} \approx 3800$~K (orange thin line) and a reduced $\chi^2 \approx 2 $ as this model severely overpredicts the $K$-band measurement. 
This discrepancy could be due the presence of broad line spectral features that either suppress the $K$-band emission or enhance the $zJ$-band measurements. 
In the latter case, if we exclude the $zJ$-band points from the fit, the continuum is well described by a hotter black-body with $T_{\rm obs} \approx 9000$~K (blue thick line). 
We examine both these models in more detail when we discuss likely progenitors and redshifts in \S\ref{sec:discussion}.




\subsection{Environment}
\label{sec:host}

\begin{table*}
	\centering
	\caption{Candidate host galaxies of GRB 210704A}
	\label{tab:host}
 \begin{tabular}{lccccccrc} 
		\hline
Label&RA&DEC&$z$&$m_r$&$R_{\rm 0}$ [\arcsec]& Offset [kpc]&$P_{\rm ch}(<R_{\rm 0})$& $R_{\rm half}$ [\arcsec]\\
\hline
    S1  & 159.0201 & 57.2163 & $2.15 \pm 0.10$ & 25.20 $\pm$ 0.20 & 0.19 $\pm$ 0.14 & 14.86 $\pm$ 11.17 & 0.02 & 0.40 \\
S2  & 159.0205 & 57.2157 & - & 24.68 $\pm$ 0.08 & 2.36 $\pm$ 0.12 & - & 0.12 & 0.40 \\
G1 & 159.0179 & 57.2243 & 0.0817 & 16.23 $\pm$ 0.00 & 29.28 $\pm$ 0.09 & 52.26 $\pm$ 0.17 & 0.005 & 2.39 \\

		\hline
	\end{tabular}
\end{table*}

The field of GRB~210704A  is shown in Figure~\ref{fig:field}. 
The GRB region shows an overdensity of low-redshift galaxies, notably
the galaxy WISEA J103604.24+571327.7 (labelled G1 in Figure~\ref{fig:field}) at $z = 0.0817$ and the galaxies' cluster 400d J1036+5713 at $z = 0.203$ \citep{{ROSAT}}. 
Two faint ($r\approx25$ AB mag) sources (labelled S1 and S2 in Figure~\ref{fig:field}) lie closer to the GRB position and are also plausible candidate host galaxies. 
Unfortunately, optical spectroscopy does not allow us to conclusively determine the redshift of GRB 210704A and use this redshift to determine the host galaxy (see \S~\ref{sec:opticalspec}). Therefore, we analyse the projected angular offsets between the galaxies and the GRB to quantify the probability of association.

We determined the positions, statistical uncertainties, and half-light radii using the pre-explosion $r-$band image from MegaPrime/MegaCam for S1 and S2, and the Gemini/GMOS-N $r-$band image obtained at $T+10.44$~d for the GRB afterglow and the other host candidates. We likewise confirm the offsets with the \textit{HST} imaging and derive the photometry for these galaxies in the \textit{HST} filters (see Table~\ref{tab:gal_phot}). To determine the offset for each galaxy we followed the methods outlined by \citet{OConnor2022}. 
We first aligned the afterglow image (Gemini) and pre-explosion image (MegaPrime) using \texttt{SExtractor} to identify point sources in each image and then \texttt{SCAMP} to determine the astrometric solution. Using 19 common point sources between the two images, we derive a relative astrometric uncertainty of $\sigma_\textrm{tie}$\,$\sim$\,$0.09\arcsec$. This astrometric uncertainty is included in the offset determination. 

Using the Equations 1 to 3 from \cite{Bloom2002} and the galaxy number counts from deep optical imaging \citep{Metcalfe2001,Kashikawa2004,McCracken2003}, the probability of finding an unrelated galaxy of magnitude $m_r$ or brighter within the vicinity of a GRB can be approximated as:
\begin{equation}
\label{pcc}
    P_{\rm ch} =1-\exp {\left(-\pi r^2_{\rm i} \times 10^{a (m_r - m_0) + b } \right )},
\end{equation}
with $a=0.36$, $b = -2.42$, and $m_0=24$ for galaxies fainter than $m_r\gtrsim19$~mag
and $a=0.56$, $b = -4.80$ and $m_0=18$ for brighter galaxies. 
The effective radius  $r_{\rm i}$ depends on the projected angular separation $R_0$ between the GRB and the galaxy and on the half-light radius $R_\textrm{half}$ of the galaxy. We take $r_{\rm i}=2R_{\rm half}$ for S1 since GRB 210704A is localized inside the detectable light, whereas for other candidates we use $r_{\rm i}=(R^2_{\rm 0}+4R^2_{\rm half})^{1/2}$ because the GRB position is well outside the light of these galaxies. In Table~\ref{tab:host}, we summarize the information on the galaxies that have $P_{\rm ch}<0.15$.

Often a galaxy is more likely to be the host than all the other ones, however, in the crowded field of GRB 210704A, two galaxies have comparably low probabilities of chance alignment with the GRB. 
The bright nearby galaxy G1, located  about 29\arcsec from the afterglow, has a low chance coincidence probability of 0.5\%.  However, the faint galaxy S1, detected close to the afterglow position, has a similar probability of 2\%. 
Hereafter, we discard S2 as a host candidate due to its high value of $P_\mathrm{ch}$.

\subsection{The Properties of the Possible Host Galaxy G1}

According to the spectrum  from the Sloan Digital Sky Survey Data Release 7, the galaxy G1 shows the signature of Ca H and K (at 3934~{\AA} and 3969~{\AA} respectively), the G-band (4304~{\AA}), Mg (5175~{\AA}) and Na (5894~{\AA}) absorption lines, all of which are consistent with an old population. Moreover, the decomposition of stellar populations and their analysis using {\sc pyPipe3D}\footnote{\url{http://ifs.astroscu.unam.mx/pyPipe3D/}}, suggest of an old dominant population. Nevertheless, the analysis confirms that a young population is also present, with about 15\% of the light corresponding to stars of less than 2~Gyr of age. 
Based on its spectral properties and observed morphology, we classify G1 as a Sa/Sb galaxy.

\subsection{The Properties of the Possible Host Galaxy S1}
\label{sec:s1}
We do not have a spectroscopic redshift for S1. Therefore, we modelled its SED (Table \ref{tab:gal_phot}) using \texttt{prospector} \citep{Johnson2019} using the methodology previously described in \citet{OConnor2021,OConnor2022}. The model parameters are the redshift $z$, total mass $M$, galaxy age $t_{\rm age}$, e-folding timescale $\tau$, extinction $A_V$, and metallicity $Z$. We use these parameters to derive the stellar mass $M_*$ and the star formation rate (SFR) as outlined in \citet{OConnor2021}. Following \citet{Mendel2014}, we apply uniform priors in log $t_{\rm age}$, log $\tau$, log $Z$, and $A_V$. We leave the redshift as a free parameter with a uniform prior between 0 and 5. We performed a fit to the data using the \texttt{DYNESTY} package \citep{dynesty}. 

\begin{figure}
\centering
 \includegraphics[width=1.0\linewidth]{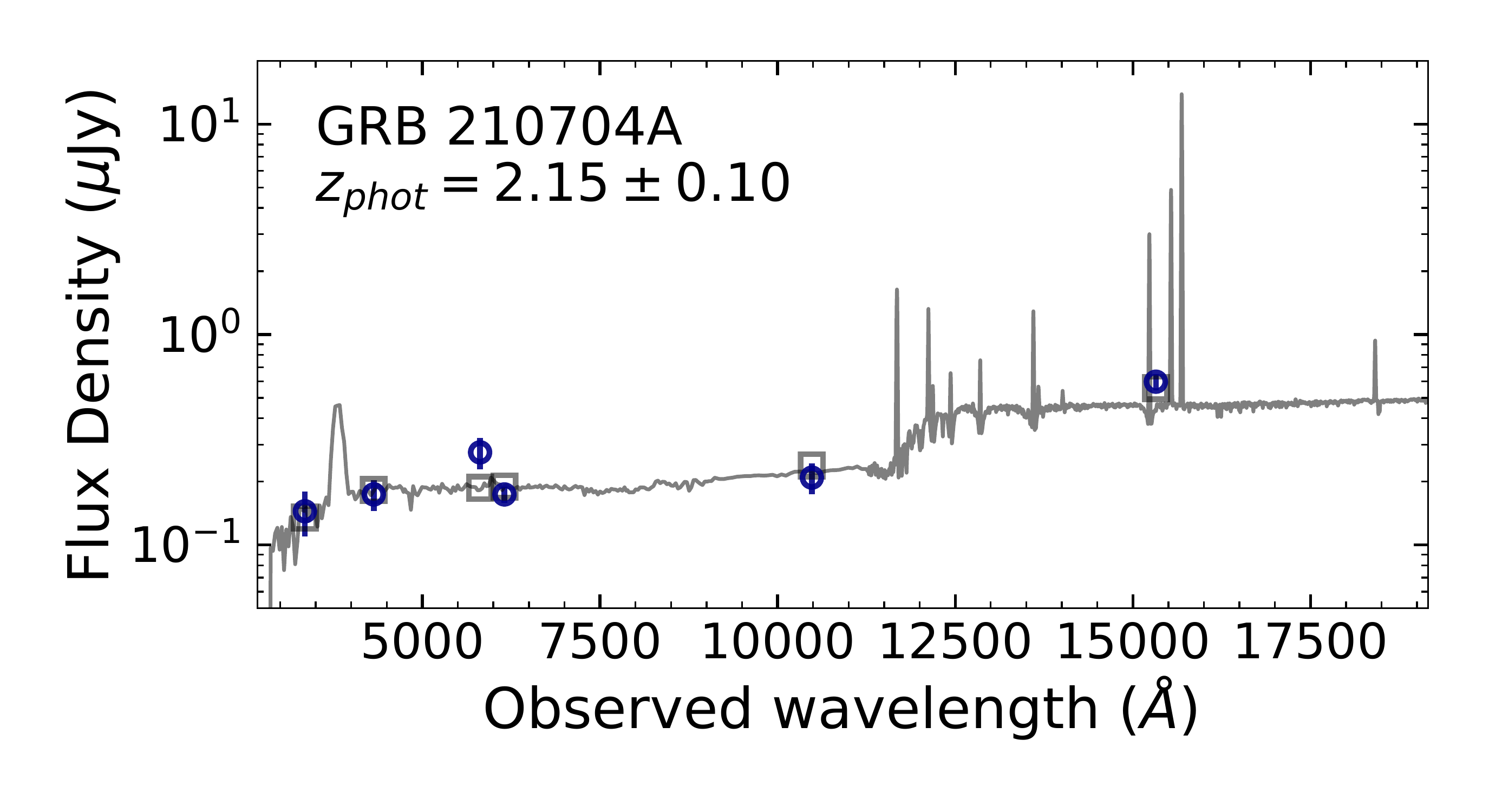}
 \vspace{-0.8cm}
 \caption{Spectral energy distribution of the galaxy S1 spatially coincident with GRB 210704A. We derive a photometric redshift of $z=2.15\pm0.10$.}
 \label{fig:galaxy_sed}
\end{figure}

We obtain a photometric redshift $z_\textrm{phot}=2.15\pm0.10$, driven by the flux increase in the 
$F160W$ filter interpreted as the 4000 \AA~break
(see Figure \ref{fig:galaxy_sed}). 
The best fit yields a stellar mass $M_*=(1-3)\times 10^{9}$ $M_\odot$, a sub-solar metallicity $Z/Z_\odot=0.3^{+0.4}_{-0.2}$, a moderate star formation rate SFR=$2.5^{+1.7}_{-0.9}$ $M_\odot$ yr$^{-1}$, a relatively young stellar population with age $t_m=0.8\pm0.5$ Gyr, and an intrinsic extinction of $A_V=0.30\pm0.15$ mag. 
These parameters are typical of GRB host galaxies in this redshift range for both long and short GRBs \citep{Palmerio2019,Dichiara2021}, although the low metallicity tends to favor the former class. 
Moreover, the galaxy's absolute magnitude of $M_{\rm r}=-21.0$ is similar to other host galaxies associated to LGRBs \citep[e.g.][]{Savaglio09}.

\subsection{A Possible Host Cluster}

The overdensity of galaxies
in this vicinity of the burst is related to the X-ray cluster 400d J1036+5713 (see Figure~\ref{fig:fieldx}). 
The cluster has an estimated mass of approximately $9\times10^{13}$~\msun~and a virial radius of $R_{500}\approx3\arcmin$. As shown in Figure~\ref{fig:fieldx}, the GRB afterglow lies about 1.5\arcmin ~from the peak of the X-ray diffuse emission, within the cluster's angular radius. 
In this case, the probabilities reported in Table~2 might overestimate the chance of a random alignment between the GRB and the galaxies in the cluster. 
To better estimate this value we consider the \textit{Swift} XRT Cluster Survey \citep{Tundo2012} and search for the number of X-ray clusters serendipitously located within 2 arcmin of a GRB position. Among over 300 long GRB fields examined in the survey, only 4 closely intercept an X-ray cluster, from which we derive a chance probability of $P_{\rm ch}\approx0.012$, comparable to the probabilities derived for S1 and G1. 
If we factor in the cluster brightness and consider only sources with $f_X \gtrsim2 \times 10^{-13}\ \mathrm{erg\,cm^{-2}\,s^{-1}}$, then the probability of a chance alignment drops to $10^{-3}$. A spurious GRB/cluster association is therefore unlikely, although not impossible when considering the large sample of more than 1500 \textit{Swift} bursts. Notably, the only other known case of an association with an X-ray cluster is the short GRB~050509B \citep{Gehrels2005,Bloom2006}.

\begin{figure}
\centering
 \includegraphics[width=.99\columnwidth]{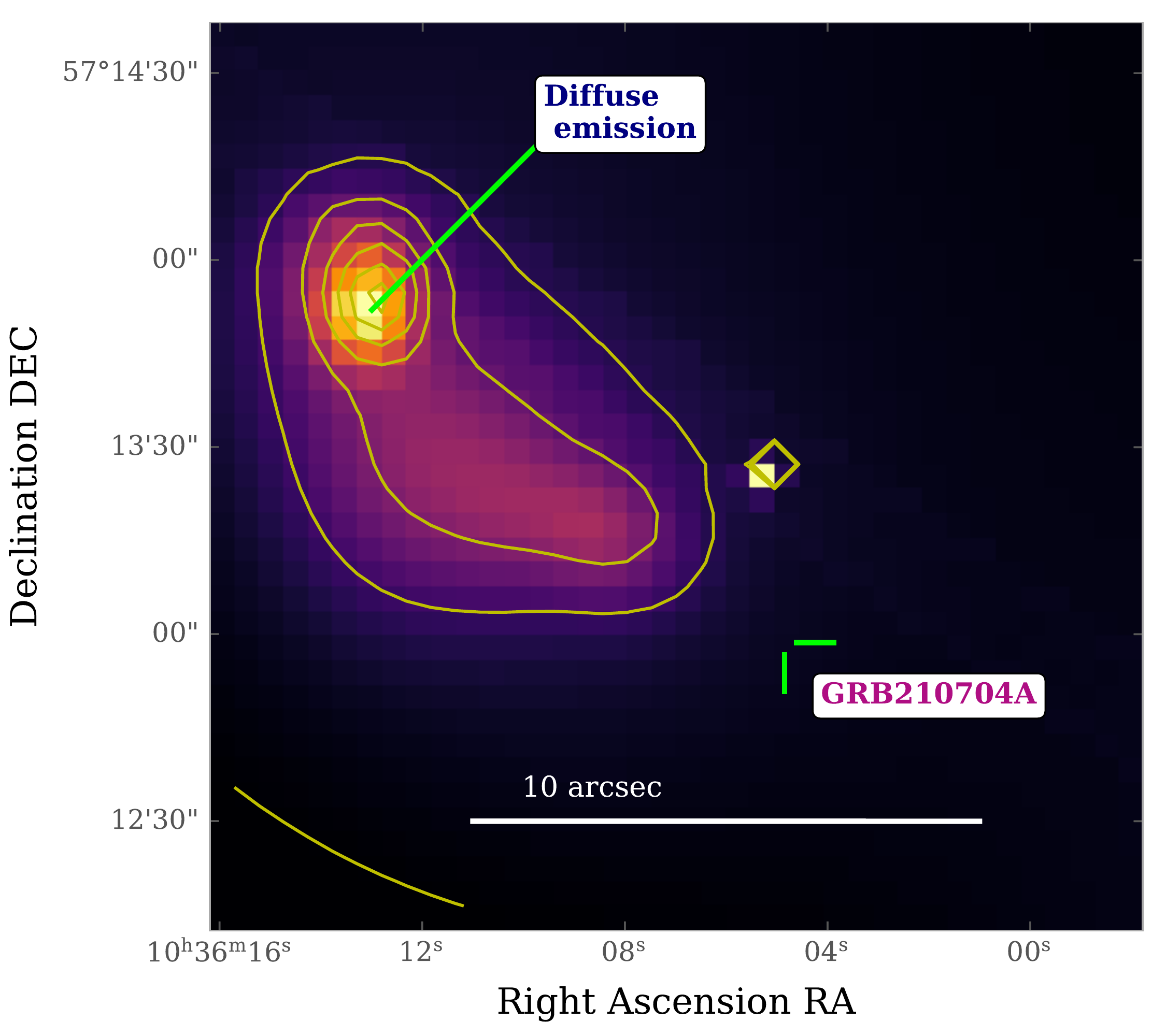}
 \caption{
 X-ray image of GRB~210704A taken by the {\it Chandra}/ACIS-S camera in the 0.5--2.0 keV energy band. The image was adaptively smoothed with a Gaussian kernel.}
  \label{fig:fieldx}
\end{figure}

\section{Discussion}
\label{sec:discussion}

In this section, we interpret and discuss the physical implications of the results obtained in section~\ref{sec:analysis}. We consider the nature of the burst (SGRB or LGRB), its redshift, and its likely environment (G1, S1 or the galaxy cluster). For distances, we assume a $\Lambda$CDM model with a $H_0=67.8$ $\mathrm{km\ s^{-1}\ Mpc^{-1}}$ \citep{Planck2014}.

\subsection{A nearby short GRB}
\label{sec:neargrb}

We first consider the possibility of a GRB hosted by the bright ($M_r\approx-21.5$) galaxy G1 (WISEA J103604.24+571327.7) at $z = 0.08168$. This is supported by the low probability of coincidence of only 0.5\% (see Table~\ref{tab:host}). 


The relatively large projected offset of about 52~kpc from G1 is an important constraint. 
The observed distribution for LGRBs does not extend beyond 10~kpc \citep{Bloom2002}, which allows us to disfavour this possibility. Furthermore, the high-energy properties of GRB~210704A do not resemble those of nearby LGRBs \citep[e.g.][]{Dichiara2021}. 

The case for a SGRB is less clear. Between 5\% and 25\% of SGRBs are observed to lie beyond 20~kpc from their host galaxies \citep{Troja2008,Berger2010,Tunnicliffe2014, OConnor2022}. Assuming an age of 1--0.1~Gyr for the GRB progenitor, the inferred offset would require an intrinsic kick velocity of $v_{\rm kick}\approx50$--$500\ \mathrm{km\ s^{-1}}$, in the range observed for Galactic NS binaries.
A SGRB at $z \sim 0.08$ would be consistent with the properties of other cosmological SGRBs (section~\ref{sec:amati}), and would not obey the spectral lag-luminosity relation of LGRBs (section~\ref{sec:lag}) as observed in the case of other nearby events \citep{Gehrels2006,Troja2022,Yang2022}. 
Therefore, based on the GRB high-energy properties and its location, a nearby SGRB remains a plausible option.

On the other hand, the observed peak luminosity and peak time of the excess emission do not match well the properties of the kilonova AT2017gfo: 
the peak brightness  would be about two magnitudes fainter ($M_r=-14.6$) than AT2017gfo ($M_r=-16.4$), and the peak time would be several days later.  It was no later than $T+12$~hours for AT2017gfo whereas for GRB 210704A we observe a maximum around $T+7$~days.

Using the one-zone model of \cite{Arnett1982}, a radioactively powered transient peaks at a time $t_\mathrm{peak}$ given by 
\begin{equation}
    t_\mathrm{peak} \approx 1.5~{\rm d~}
    \bigg(\frac{M_\mathrm{ej}}{0.01 M_\odot}\bigg)^{0.5}
    \bigg(\frac{\kappa}{1\ \mathrm{cm^2\ g^{-1}}}\bigg)^{0.5}
    \bigg(\frac{v_\mathrm{ej}}{0.1 c}\bigg)^{-0.5}
    \label{eq:arnett}
\end{equation}
where $M_{\rm ej}$ is the ejecta mass,  
$v_{\rm ej}$ its velocity, and $\kappa$ its opacity. 
To explain a peak of about 7~days, one needs to either increase the mass to of order 0.1{\msun} and/or increase the opacity. As a result, the spectral peak would shift to redder wavelengths than the ones observed in this case. 
Alternatively, a delayed peak can be explained by a low expansion velocity, $v\approx0.01 c$. This latter solution
is consistent with the simple blackbody model for the optical/nIR excess, which, at a redshift of 0.08, implies temperatures of 4000-10000~K and expansion velocities of 20000-2000 km s$^{-1}$, slower than typical merger ejecta. Therefore, this solution does not seem consistent with a merger progenitor.

We also compared our case to the library of simulated kilonova light curves  by \citet{Wollaeger2021}, incorporating a broad range of ejecta compositions, morphologies and viewing angles. 
We find that, although the luminosities of the kilonova models are in agreement with the observations for $z \lesssim 0.4$,  the optical peak can only range from a few hours up to a couple of days after the explosion, whereas we observe a peak at $T$+7~d.
We note that the mismatch between the observed peak and the kilonova timescale applies to any SGRB up to $z\approx1$, and therefore we can rule out any typical kilonova in this range. 

\subsection{A distant long GRB}
\label{sec:near-lgrb-s1}

We next consider the possibility that GRB 210704A is associated with the faint source S1. Intuitively, this may appear as the most likely association given the positional coincidence between S1 and the GRB. However,  the probability of chance coincidence is 2\% (Table~\ref{tab:host}), larger than G1.

The high-redshift derived from the photometric fit in \S~\ref{sec:s1} is consistent with the weak absorption feature in the afterglow optical spectrum (section \ref{sec:opticalspec}). This provides us with tantalizing evidence that GRB 210704A was indeed located at $z\approx2.3$.
At this distance, its rest-frame duration would be $\lesssim$1.5~s.
However, its other high-energy properties such as spectral lag (section~\ref{sec:lag}), peak energy, and energetics (section~\ref{sec:amati}), would fit within the distribution of typical LGRBs (see Figure~\ref{fig:amati-lag}).

In this high-redshift scenario, the case of GRB 210704A is reminiscent of other bursts, such as GRB 090426A \citep[][]{Antonelli09} and GRB 200826A \citep[][]{Ahumada2021, Zhang2021}, characterized by a short duration of the gamma-ray emission despite likely having collapsing massive stars as progenitors. 
Based on the observed BATSE distribution of GRB durations \citep[][]{Kouveliotou93}, a small fraction of LGRBs ($\lesssim$1\%) are expected to last less than 2~s. It is therefore not surprising to find some of these examples within the sample of over 1,000 bursts discovered by \textit{Swift}.

However, in the case of GRB 210704A, the observed optical bump would be challenging to explain with standard SN models.
At a redshift of 2.3, the simple blackbody model adopted for the optical/nIR excess implies a
temperature of 30000~K and a sub-relativistic velocity of $\sim 0.2$ c. The second solution of a colder black body peaking in the observed $J$-band requires a superluminal speed, and is thus considered unphysical.

The observed magnitude of the excess of $r\approx23$ AB at $T+7$~d implies an absolute magnitude of $M_{UV}\approx-23.2$ at $T+2.3$~d (rest-frame) (after subtracting the afterglow contribution), brighter than most known SNe \citep[e.g.][]{Pastorello2010,Richardson2014} and AT2018cow-like events \citep[e.g.][]{Perley2019} (see Figure~\ref{fig:cow}). The only event whose brightness in UV is comparable with the excess of GRB 210704A is the Dougie tidal disruption event (TDE) candidate \citep{Vinko2015}. However, 
known TDEs evolve on slower timescales than the excess of GRB210704A. 

Mechanisms capable of producing such a high peak brightness and rapid rise time are rare.
For instance, the interaction between the SN shock wave and a dense environment, or the continued energy injection from a long-lived central engine are also expected to act on longer timescales than the ones observed here \citep[e.g.][]{Cartier2021} . 
\citet{Fryer2007} suggested that, for the most energetic explosions ($\gtrsim$10$^{52}$ erg), a shock breakout propagating through a dense stellar wind could produce strong emission in excess of 10$^{44}$ erg s$^{-1}$, peaking 
in the UV on timescales of a few days after the burst. This emission becomes dominant when the Ni$^{56}$ yield is low
as expected if the star directly collapses to a black hole. 
Although expected theoretically, the fraction of failed SN, in which most of the star collapses to a black hole, is not well constrained observationally \citep[see e.g.][]{Adams2017}.


\begin{figure}
\centering
 \includegraphics[width=1.0\columnwidth]{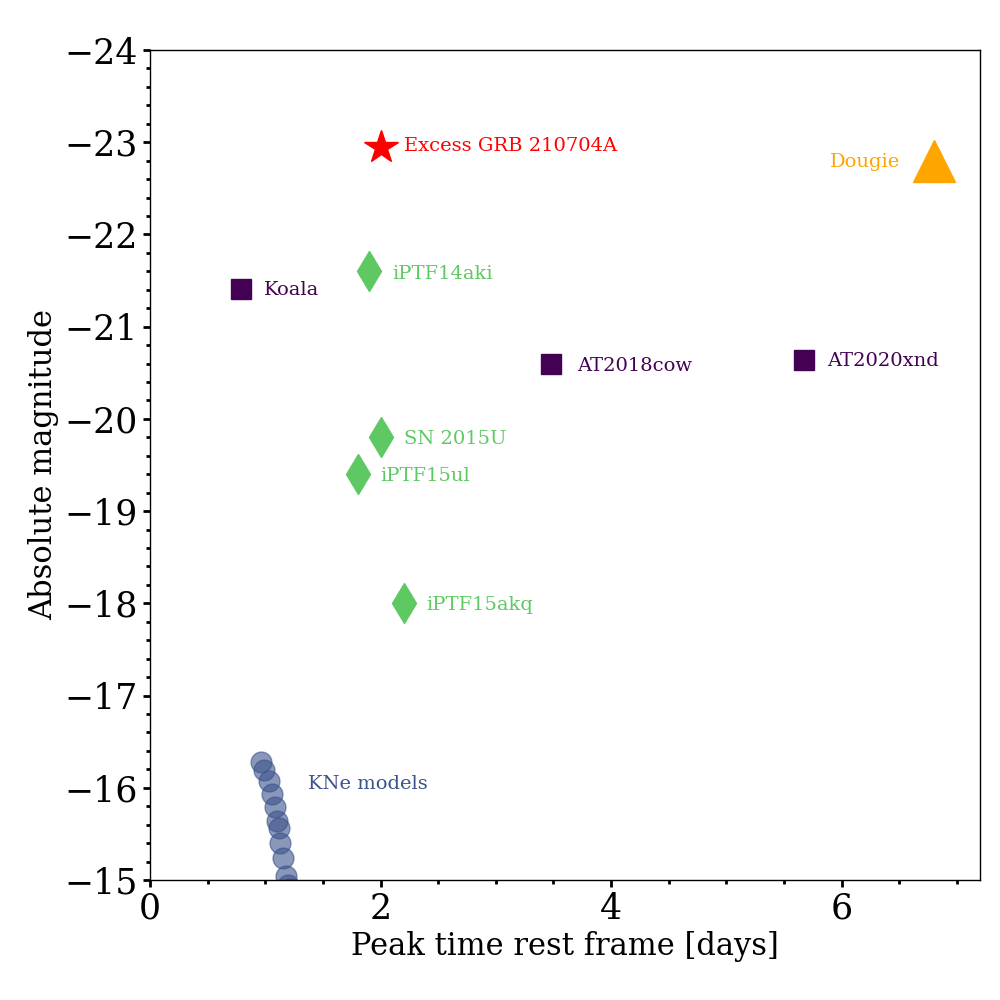}
 \caption{Phase space of luminosity and peak time of different transients.
 Assuming a redshift of 2.34 for GRB210704A, the observed optical excess  (red star) implies highly luminous 
 blue emission (at a rest-frame effective length $\lambda_\mathrm{rest}=191$~nm). 
 This is compared with fast blue optical transients: AT2018cow ($\lambda_\mathrm{rest}=172$~nm) \citep{Perley2019}, KOALA ($\lambda_\mathrm{rest}=382$~nm) \citep{Ho2020} and  AT2020xnd ($\lambda_\mathrm{rest}=288$~nm) \citep{Perley2021} (purple squares), simulated kilonova lightcurves by \citet{Wollaeger2021} ($\lambda_\mathrm{rest}=470$~nm) (blue points), Type Ibn Supernovae \citep{Hosseinzadeh2017} SN2015U ($\lambda_\mathrm{rest}=359$~nm), iPTF15ul ($\lambda_\mathrm{rest}=342$~nm), iPTF14aki ($\lambda_\mathrm{rest}=342$~nm), iPTF15akq ($\lambda_\mathrm{rest}=328$~nm) (green diamonds) and Dougie, a tidal disruption event candidate \citep{Vinko2015} ($\lambda_\mathrm{rest}=543$~nm) (orange triangle).}
 \label{fig:cow}
\end{figure}

\subsection{A nearby peculiar GRB}
\label{sec:cluster}

A final possibility is that GRB 210704A is a peculiar GRB associated with a member of the galaxy cluster at $z = 0.203$. 
\cite{Zemp2009} showed that progenitor systems formed within a cluster are likely to be retained within the potential well of its halo, regardless of the intrinsic kick velocity of the progenitor system. However, due to the high merger activity in such crowded environment, the GRB may happen far from its true birthplace. As a result, its host galaxy cannot be confidently identified based on its spatial proximity. 

A cluster environment, although rarely associated with a GRB, has been discussed in the case of both SGRBs and LGRBs. 
The connection with SGRBs naturally arises as NS binaries with long merger timescales are expected to track the stellar mass density and reside in evolved stellar environments. 
However, as discussed in section~\ref{sec:neargrb}, a SGRB produced by a typical NS merger would not produce a kilonova compatible with the luminosity and late-onset of the optical excess.

In the case of LGRBs, their typical progenitors, young massive stars \citep{Woosley2006}, are not found in clusters, and a merger-driven explosion with unusual progenitors appears to be the most likely explanation. For example, black hole and neutron star systems (BH-NS), black hole and white dwarf (BH-WD), white dwarf and neutron star (WD-NS) or WD-WD  encounters have been considered \citep{Fryer1998,Page2006,King2007, Yang2022}. 

An exotic progenitor system could explain the unusual high-energy properties of this event and why they do not fit within the standard relations of typical LGRBs for $z\sim0.2$. In addition, we note that the only other LGRB possibly associated with a galaxy cluster is GRB~050911 \citep{Page2006}, whose prompt emission was a short duration ($\sim$1~s) peak followed by a weak 16~s long tail and as such resembles the light curve of GRB~210704A. A LGRB produced by an exotic merger system could also give rise to a peculiar optical transient, although in this case, predictions are less secure. 

Assuming that the optical excess is radioactively-powered, we consider the case of ejecta with pure nickel abundance. This is justified by the rapid evolution of the optical emission, comparable to the $^{56}$Ni half-life timescale  of about 6~d. 
At a redshift of $z \approx 0.2$, the optical bump would peak at an absolute magnitude of $M_r\approx-16.5$, corresponding to a luminosity of about $10^{42}$~erg~s$^{-1}$. The peak bolometric luminosity can be related to the ejecta mass $M_{ej}$ using Arnett's law, $L_{pk} \approx \epsilon_{\rm Ni}(t_{pk}) M_{ej}$, where $\epsilon_{\rm Ni}(t_{pk}) \approx 5\times10^{43}$~erg~s$^{-1}$ is the energy generation rate at 6~d (rest-frame). 
The derived mass $M_{ej}\approx0.1$ \msun~is much smaller than in a typical SN, and indicative of a weaker explosion. 
Based on the black-body fit of the optical excess, we derived a temperature of 4500--10000~ K and ejecta velocity $v_{\rm ej}\approx0.2-0.02 c$. 
Therefore, the basic properties of the  optical excess, such as its luminosity and fast timescales, could be reproduced by an explosion ejecting a low-mass shell of fast-moving material, composed mainly of $^{56}$Ni. 
At first order, this is consistent with the predictions of an  accretion-induced collapse (AIC) of a WD \citep{SharonKushnir2020}, 
driven for example from a WD-WD encounter \citep{LyutikovToonen2019,Rueda2018}.

\section{Summary}
\label{sec:summary}

We have presented broadband observations of GRB 210704A, its afterglow, and environment.
After its discovery, the burst was initially classified as a SGRB.
Our analysis of the high-energy prompt emission shows that the GRB duration is affected by instrumental selection effects, and that the GRB lies at the intersection between SGRBs and LGRBs.
We place the GRB in the context of the Amati and spectral lag correlations, and show that these can provide information on the nature of the GRB at different possible redshifts.

An additional peculiar feature of this GRB is an optical/nIR excess, observed to peak at about $T+7$ days (observer's frame)
at a magnitude $r\approx23.2$ AB. 
We identified this excess by comparing our optical and nIR photometry with a simple afterglow model derived from the X-ray data.


We consider three possible host environments and distance scales: the galaxy G1 at $z \approx 0.08$, the galaxy S1 at $z \approx 2.3$, and a galaxy's cluster at $z\approx0.2$. We also discuss three different progenitors: a standard SGRB with a kilonova, a standard LGRB with a SN, and an exotic LGRB, perhaps an accretion-induced collapse of a WD \citep{LyutikovToonen2019,Rueda2018,SharonKushnir2020}.

Neither of the standard explanations is entirely satisfactory. A SGRB followed by a kilonova, perhaps associated with G1, explains the short duration of the prompt gamma-ray emission, its other high-energy properties, and the large distance from the host galaxy. However, the color and timescales of the observed optical excess imply an expansion velocity not typical for the ejecta of a compact binary merger. 

On the other hand, a LGRB associated with the galaxy S1 remains consistent with all the high-energy properties of the event but is challenged by the extreme luminosity of the optical excess. No known transient matches the high luminosity, blue color, and rapid timescales implied by the observations.  

Finally, we considered an exotic progenitor, involving the merger of a WD with another compact objects (either a BH, a NS, or another WD), in the cluster at $z \approx 0.2$. 
Such stellar encounters can explain GRB durations longer than the canonical 2 seconds and could be followed by fast-evolving optical transients consistent with the observed excess. This is the only scenario that matches all the observed properties of GRB210704A although, admittedly, is also the least constrained one.

The difficulty we have had in identifying the progenitor of GRB 210704A highlights the limitations of the traditional GRB dichotomy in long/short, collapsars/mergers events. Although valid for the majority of bursts, it is not sufficient to describe the GRB population in its entirety. The large \textit{Swift} sample of well-localized bursts has been instrumental to identify these oddball GRBs and start exploring their possible origins. In the next few years, the arrival of the Vera C. Rubin Observatory will undoubtedly enlarge the sample of exotic transient events and perhaps help us find analogues to GRB 210704A.


\section*{Research Data Policy}
The data underlying this article will be shared on reasonable request to the corresponding author.\\

\section*{ACKNOWLEDGEMENTS}

We thank the staff of GTC and Gemini for scheduling and executing the observations included in this work, especially Antonio Cabrera and  David Garcia.

This project has received funding from the European Research Council (ERC) under the European Union’s Horizon 2020 research and innovation programme, grant  101002761 (BHianca; PI: Troja).

Some of the data used in this paper were acquired with the RATIR instrument, funded by the University of California and NASA Goddard Space Flight Center, and the 1.5-meter Harold L. Johnson telescope at the Observatorio Astronómico Nacional on the Sierra de San Pedro Mártir, operated and maintained by the Observatorio Astronómico Nacional and the Instituto de Astronomía of the Universidad Nacional Autónoma de México. Operations are partially funded by the Universidad Nacional Autónoma de México (DGAPA/PAPIIT IG100414, IT102715, AG100317, IN109418, IG100820, and IN105921). We acknowledge the contribution of Leonid Georgiev and Neil Gehrels to the development of RATIR.

Some of the data used in this paper were acquired with the DDOTI instrument at the Observatorio Astronómico Nacional on the Sierra de San Pedro Mártir. DDOTI is partially funded by CONACyT (LN 232649, LN 260369, LN 271117, and 277901), the Universidad Nacional Autónoma de México (CIC and DGAPA/PAPIIT IG100414, IT102715, AG100317, IN109418, and IN105921), the NASA Goddard Space Flight Center and is partially funded by the University of Maryland (NNX17AK54G). DDOTI is operated and maintained by the Observatorio Astronómico Nacional and the Instituto de Astronomía of the Universidad Nacional Autónoma de México. We acknowledge the contribution of Neil Gehrels to the development of DDOTI.

We thank the staff of the Observatorio Astronómico Nacional.

We thank Ori Fox, Antonio Castellanos-Ramírez, Yuri Cavecchi and Sebastián F. Sánchez for their useful comments. We also thank Rubén Sánchez Ramírez and Aishwarya Thakur for their support in the reduction of GTC data.

We acknowledge support from the DGAPA/UNAM IG100820 and IN105921. RLB acknowledges support from CONACyT and DGAPA postdoctoral fellowships.

M.I. and G.S.H.P. acknowledge the support from the National Research Foundation grants No. 2020R1A2C3011091 and No. 2021M3F7A1084525, and the R\&D program (Project No. 2022-1-860-03) of the Korea Astronomy and Space Science Institute.

This research is based in part on data collected at the Subaru Telescope, which is operated by the National Astronomical Observatory of Japan. We are honored and grateful for the opportunity of observing the Universe from Maunakea, which has the cultural, historical, and natural significance in Hawaii. 

This work is partly based on data obtained with the Gran Telescopio Canarias (GTC), installed in the Spanish Observatorio del Roque de los Muchachos of the Instituto de Astrofísica de Canarias, in the island of La Palma, and with the instrument OSIRIS, built by a Consortium led by the Instituto de Astrofísica de Canarias in collaboration with the Instituto de Astronomía of the Universidad Autónoma de México. OSIRIS was funded by GRANTECAN and the National Plan of Astronomy and Astrophysics of the Spanish Government.

The scientific results reported in this article are based in part on observations made by the Chandra X-ray Observatory through Director's Discretionary Time (ObsID: 25093; PI: Troja). 

This work made use of data supplied by the UK Swift Science Data Centre at the University of Leicester. 

RLB dedicates this work to Cicerón. Thanks for sharing your life with me. I will always love you.






\bsp	
\label{lastpage}
\end{document}